\documentclass[12pt]{phb-proc4-auth}

\usepackage{graphicx}
\usepackage{amssymb}

\begin{document}

\begin{frontmatter}

\journal{SNS'2004}

\title{Resonant Enhancement of Electronic Raman Scattering}

\author[UA]{A.~M. Shvaika\corauthref{1}}
\author[UA]{O. Vorobyov}
\author[US]{J.~K.~Freericks}
\author[CA]{T.~P.~Devereaux}

\address[UA]{Institute for Condensed Matter Physics, Lviv, Ukraine} 
\address[US]{Department of Physics, Georgetown University, Washington, DC, 
20057 USA}
\address[CA]{Department of Physics, University of Waterloo, Ontario, N2L 3GI
Canada}

\corauth[1]{Corresponding Author: Institute for Condensed Matter Physics of 
the National
Academy of Sciences of Ukraine, 1 Svientsitskii Street, 79011 Lviv, Ukraine,
Phone: +380-322-761054, Fax: +380-322-761158,
Email: ashv@icmp.lviv.ua}

\begin{abstract}
We present an exact solution for electronic Raman scattering in a single-band, strongly correlated material, including nonresonant, resonant and mixed contributions. Results are derived for the spinless Falicov-Kimball model, employing dynamical mean field theory; this system can be tuned through a Mott metal-insulator transition. 
\end{abstract}

\begin{keyword}
Resonant Raman scattering \sep Falicov-Kimball model
\end{keyword}

\end{frontmatter}

Electronic Raman scattering is an important probe of electronic excitations
in materials.  It has been used to examine different kinds of charge and
spin excitations in a variety of different materials, ranging from Kondo
insulators~\cite{Pap1,Pap2}, to high temperature 
superconductors~\cite{resonance,blumberg}, to collossal magnetoresistance
materials~\cite{Pap3}.  Inelastic light scattering involves 
contributions from scattering
processes that depend on the incident photon frequency (so-called mixed and
resonant contributions) and processes that are independent of the incident
photon frequency (so-called nonresonant contributions). There has been much
theoretical work on this problem.  In the strong-coupling regime, a 
perturbative approach has been used, and has illustrated a number of important
features of resonant scattering processes~\cite{Shastry,Chubukov}.  The 
nonresonant case has also
been examined, and an exact solution for correlated systems (in large
spatial dimensions) is available
for both the Falicov-Kimball~\cite{paper1} and Hubbard~\cite{paper2} models.  
Here we concentrate on an
exact solution of the full problem for the Falicov-Kimball
model including all resonant and mixed effects.

The Falicov-Kimball model Hamiltonian~\cite{falicov_kimball} is (at half filling)
\begin{eqnarray}
\mathcal{H}=-\frac{t^*}{\sqrt{d}}\sum_{\langle ij \rangle} c^\dagger_ic_j
+U\sum_i \left(c^\dagger_ic_i-\frac12\right)\left(w_i-\frac12\right)
\label{eq: ham_def}
\end{eqnarray}
and includes two kinds of particles: conduction electrons, which are mobile, 
and localized electrons which are immobile.  
Here $c^\dagger_i$ ($c_i$) creates (destroys) a conduction electron at
site $i$, $w_i$ is the localized electron number at
site $i$, $U$ is the on-site Coulomb interaction between the
electrons, and $t^*$ is the hopping integral (which we use as our energy unit).
The symbol $d$ is the spatial dimension, and
$\langle i j\rangle$ denotes a sum over all nearest neighbor pairs
(we work on a hypercubic lattice).
The model is exactly solvable with dynamical mean field 
theory when $d\rightarrow\infty$~\cite{metzner_vollhardt,brandt_mielsch}
(see ~\cite{freericks_review} for a review).

Shastry and Shraiman~\cite{Shastry}
derived an explicit formula for inelastic light scattering
that involves the matrix elements of the electronic vector potential
for light with the many-body states of the correlated system. The expression for the Raman response is
\begin{eqnarray}
R(\Omega)&=&2\pi \sum_{i,f} \exp(-\beta\varepsilon_i)
    \delta(\varepsilon_f - \varepsilon_i - \Omega)\nonumber
    \\
&\times& \left| \frac{hc^2}{V\sqrt{\omega_i\omega_o}}e_\alpha^i e_\beta^o
    \left\langle f \left| \hat M^{\alpha\beta}\right| i \right\rangle
    \right|^2 \big /\mathcal{Z}
\label{eq: raman1}
\end{eqnarray}
for the scattering of electrons by optical photons
(the repeated indices $\alpha$ and $\beta$ are summed over).
Here $\varepsilon_{i(f)}$
refer to the initial (final) eigenstates describing the ``electronic matter'', 
$\mathcal{Z}$ is the partition function, and  
\begin{eqnarray}\label{M_oper}
    &&\left\langle f \left| \hat M^{\alpha\beta} \right| i \right\rangle
    =
    \left\langle f \left| \gamma_{\alpha,\beta} \right| i \right\rangle\nonumber
    \\
    &&+ \sum_l \left(
    \frac{\left\langle f \left| j_{\beta} \right| l \right\rangle
    \left\langle l \left| j_{\alpha} \right| i \right\rangle}
    {\varepsilon_l - \varepsilon_i - \omega_i}
+
    \frac{\left\langle f \left| j_{\alpha} \right| l \right\rangle
    \left\langle l \left| j_{\beta} \right| i \right\rangle}
    {\varepsilon_l - \varepsilon_i + \omega_o}
    \right)
\end{eqnarray}
is the scattering operator constructed by the current $j_\alpha=\sum_{\bf k}
\partial\varepsilon({\bf k})/
\partial k_\alpha c^\dagger_{\bf k}c_{\bf k}$, and stress-tensor $\gamma_{\alpha\beta}=
\sum_{\bf k}\partial^2\varepsilon({\bf k})/\partial k_\alpha\partial k_\beta
c^\dagger_{\bf k}c_{\bf k}$ operators, with $\varepsilon({\bf k})$ the band structure and $c_{\bf k}$ the destruction operator for an electron with momentum ${\bf k}$.  

Instead of calculating the matrix elements in Eq.~(\ref{M_oper}), we use a diagrammatic technique to calculate all contributions to the Raman response function $\chi(\Omega)$, which is defined by
\begin{equation}\label{Rfin}
    R(\Omega) = \frac{2\pi h^2 c^4}{V^2\omega_i\omega_o}\;
    \frac{\chi(\Omega)}{1-\exp(-\beta\Omega)} .
\end{equation}
Our calculations include effects from nonresonant diagrams, from resonant diagrams, and from so-called mixed diagrams.  
We have evaluated these 
expressions for the Stokes Raman response, with an incident photon
frequency $\omega_i$, an outgoing photon frequency $\omega_o$, and a
transfered photon frequency $\Omega=\omega_i-\omega_o$.  The procedure is
complicated, and involves first computing the response functions on the 
imaginary time axis, then Fourier transforming to imaginary frequencies,
and finally performing an analytic continuation to the real axis~\cite{details}.

We analyze three different symmetries for the
incident and outgoing light.
The $A_{\mathrm{1g}}$ symmetry has the full symmetry of the
lattice and is measured by taking the initial and
final polarizations to be $\textbf{e}^i=\textbf{e}^o=(1,1,1,...)$ 
(we assume nearest-neighbor hopping only).
The $B_{\mathrm{1g}}$ symmetry is a $d$-wave-like symmetry  that involves 
crossed polarizers: $\textbf{e}^i=(1,1,1,...)$ and
$\textbf{e}^o=(-1,1,-1,1,...)$.  Finally,
the $B_{\mathrm{2g}}$ symmetry is another $d$-wave symmetry rotated by
45 degrees; with $\textbf{e}^i=(1,0,1,0,...)$ and $\textbf{e}^o=(0,1,0,1,...)$. 


The total Raman response function is the sum of the nonresonant, mixed, and resonant contributions and has a complicated form. 
It turns out that the $A_{\mathrm{1g}}$ sector has contributions
from nonresonant, mixed, and resonant Raman scattering, the $B_{\mathrm{1g}}$
sector has contributions from nonresonant and resonant Raman scattering only,
and the $B_{\mathrm{2g}}$ sector is purely resonant~\cite{paper1}.
It is educational to consider the contributions of the bare diagrams, which can be summed up and rewritten in the following form:
\begin{eqnarray}\label{chi_bare}
    &&\chi_{\rm bare}(\Omega)=\frac1N\sum_{\bf k}\int^{+\infty}_{-\infty} d\omega
    [f(\omega)-f(\omega+\Omega)] 
    \\
    \nonumber
    &&\qquad\times A_{\bf k}(\omega) A_{\bf k}(\omega+\Omega)
    \\
    \nonumber
    &&\times\left|\gamma_{\bf k}+v_{\bf k}^{i}v_{\bf k}^{o}
    \left[G_{\bf k}(\omega+\omega_i+i\delta)+G_{\bf k}(\omega-\omega_o-i\delta)\right]\right|^2,
\end{eqnarray}
where $\gamma_{\bf k}=\sum_{\alpha,\beta} e_\alpha^i
\frac{\partial^2\epsilon_{\bf k}}{\partial k_\alpha\partial k_\beta}
e_\beta^o$, $v_{\bf k}^{i,o}=\sum_{\alpha} e_\alpha^{i,o}
\frac{\partial\epsilon_{\bf k}}{\partial k_\alpha}$, $A_{\bf
k}(\omega)=-\frac1\pi\mathop{\mathrm{Im}} G_{\bf k}(\omega-i\delta)$, 
$G_{\bf k}(\omega)$ is the mo\-men\-t\-um-dependent single-electron Green's function, and $f(\omega)=1/[1+\exp (\beta\omega)]$  is the Fermi distribution
function.

In general, the bare response function in Eq.~(\ref{chi_bare}) is a function of 
the frequency shift $\Omega=\omega_i-\omega_o$, of the incoming photon 
frequency $\omega_i$ and the outgoing photon frequency $\omega_o$; it
can be enhanced when one or both of the denominators of the Green's functions of the effective vertex (expression on the last line) are resonant. In the latter case, we have a so-called ``double'' or ``multiple
resonance'' \cite{MartinFalicov}. The full response function also includes 
vertex renormalizations. But the total (reducible) charge vertex for the Falicov-Kimball model does not diverge, and hence it 
does not introduce any additional energy denominators or ``resonances''; it only leads to a renormalization of the total Raman response.


The Falicov-Kimball model on a $d=\infty$ hypercubic lattice has a Mott transition
into a pseudogap-like phase at half filling when $U=\sqrt{2}$. 
We examine the system on the insulating side of the Mott transition at $U=3$.

\begin{figure}
\centering
\includegraphics[height=0.99\columnwidth,angle=-90]{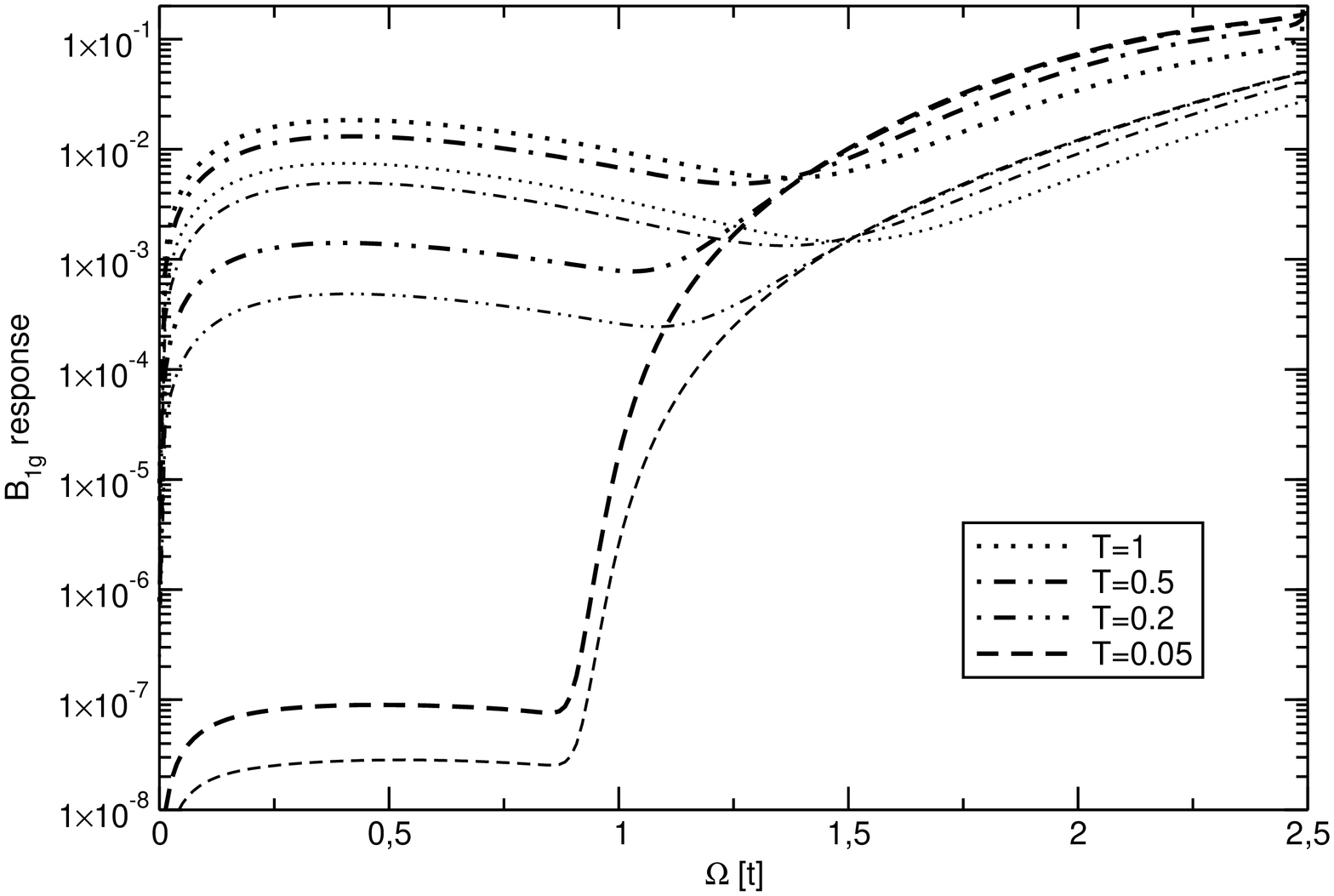}\\
\includegraphics[height=0.99\columnwidth,angle=-90]{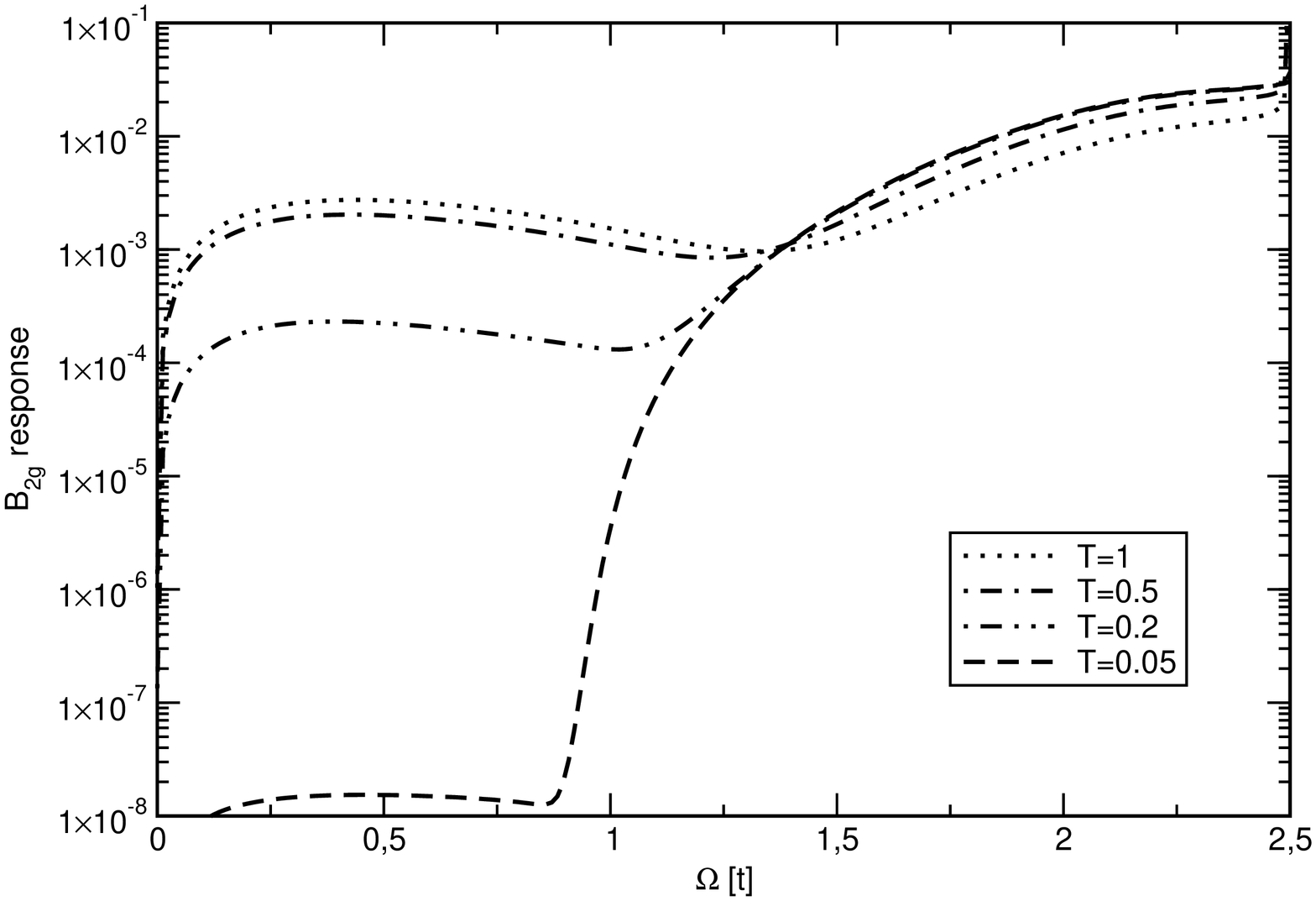}\\
\includegraphics[height=0.99\columnwidth,angle=-90]{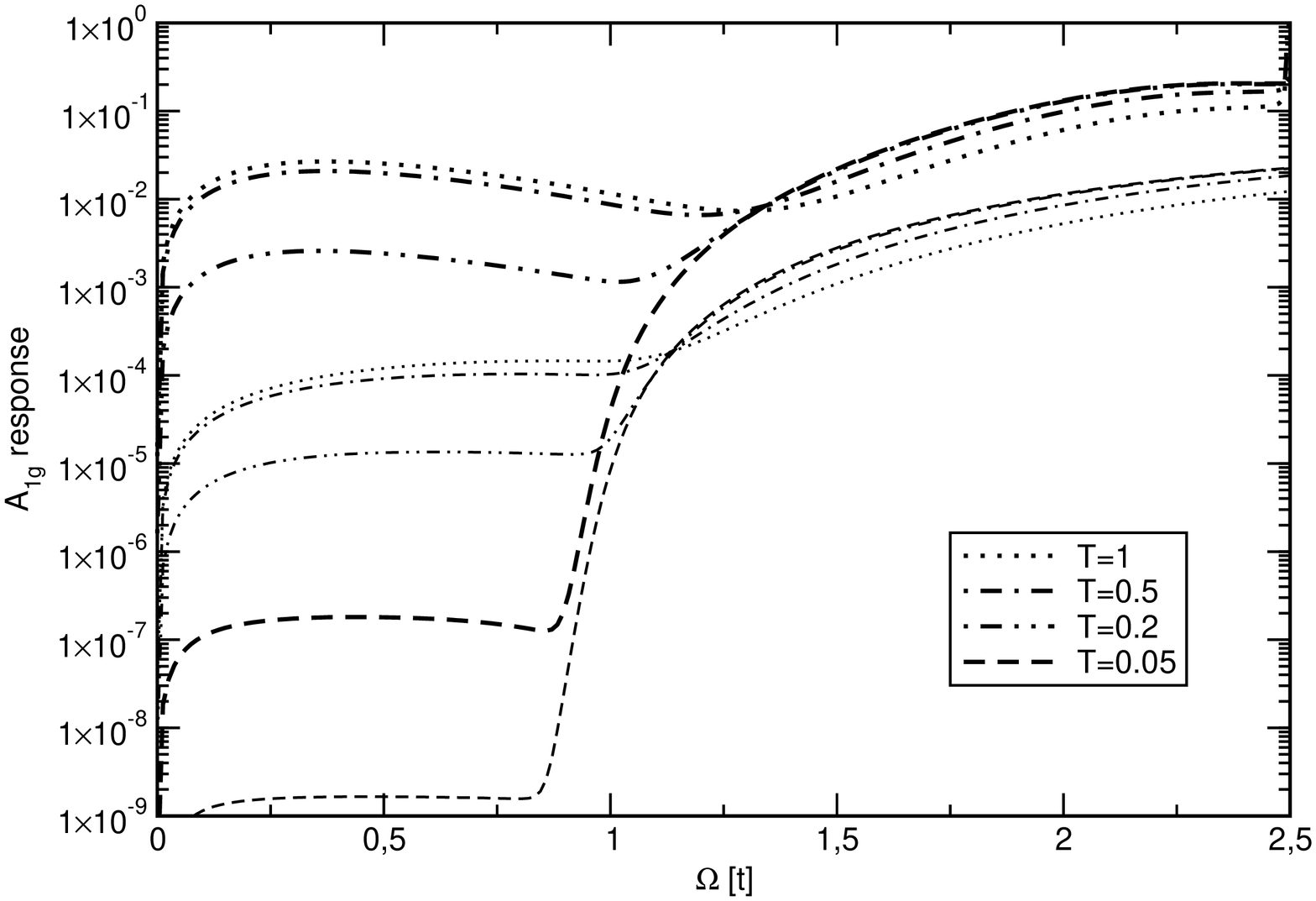}
\caption{Isosbestic behavior of the resonant Raman response for $U=3$ and $\omega_i=2.5$ (thick lines) and $\omega_i=\infty$ (nonresonant response, thin lines). Different lines correspond to different temperatures $T=1$, $0.5$, $0.2$, $0.05$. Note that the $\omega_i=2.5$ curves all cross at two isosbestic points: one close to $U/2$ and another close to $\omega_i$. \label{fig:raman_iso1}}
\end{figure}

\begin{figure}[htbf]
\centering
\includegraphics[height=0.99\columnwidth,angle=-90]{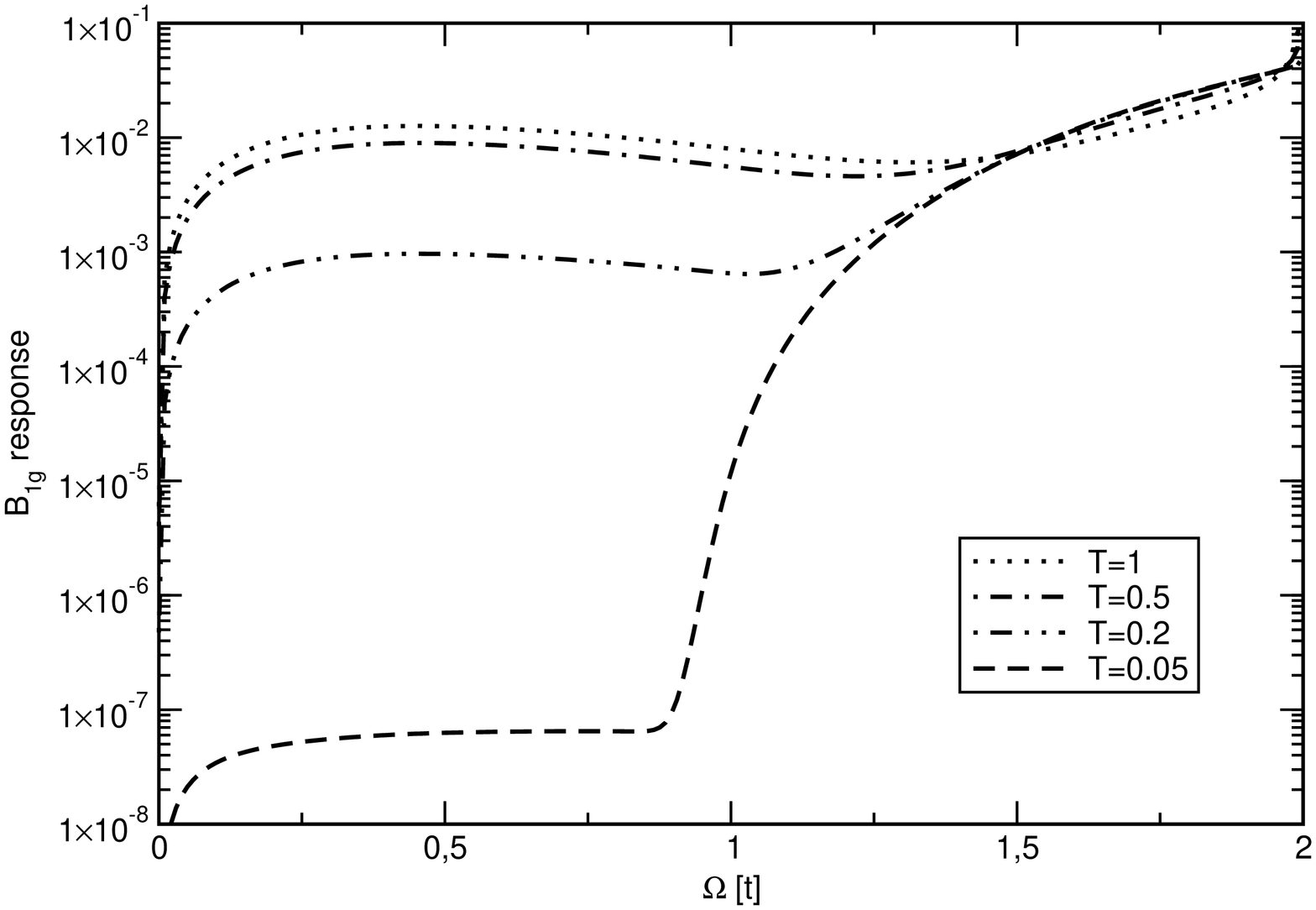}\\
\includegraphics[height=0.99\columnwidth,angle=-90]{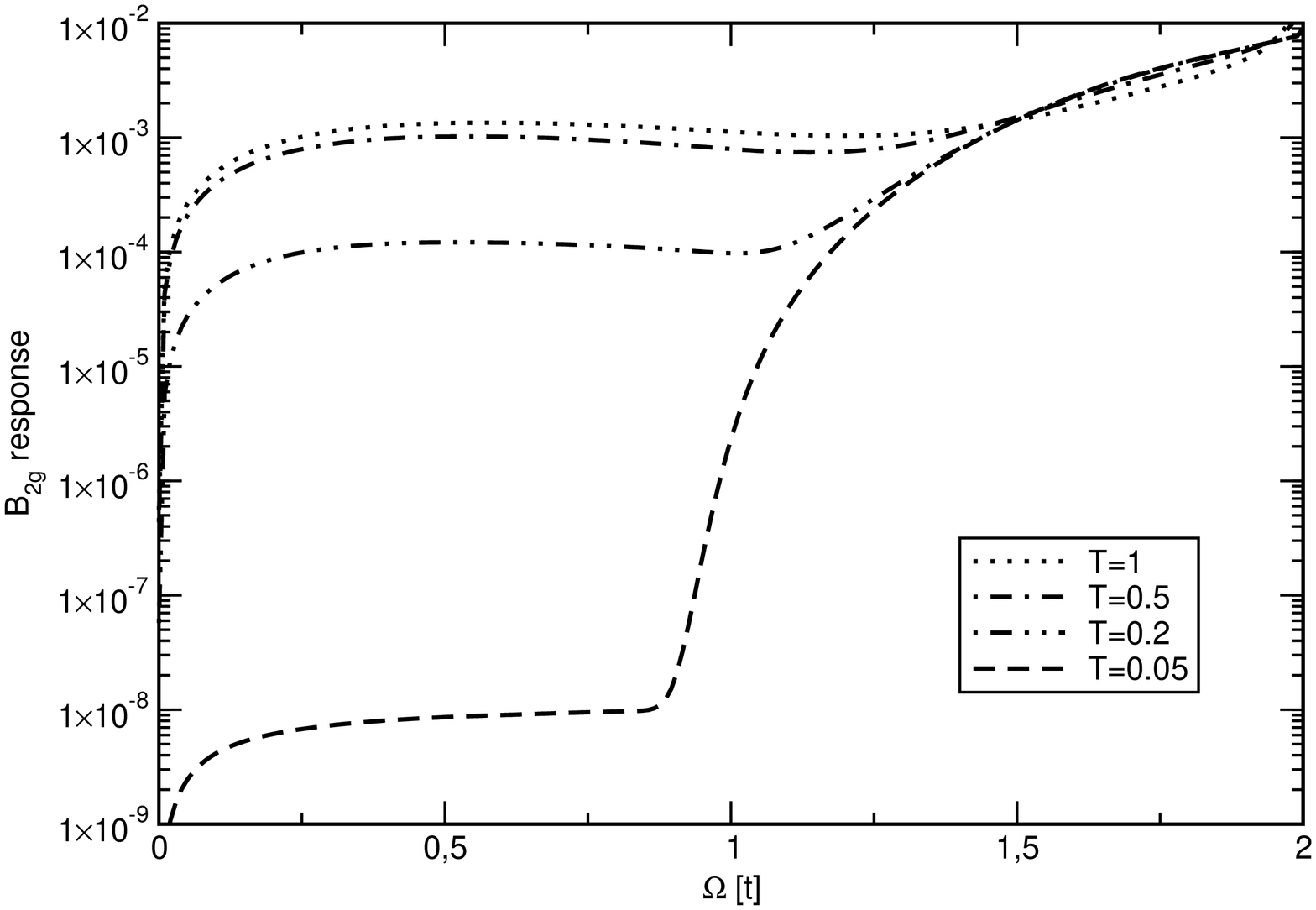}\\
\includegraphics[height=0.99\columnwidth,angle=-90]{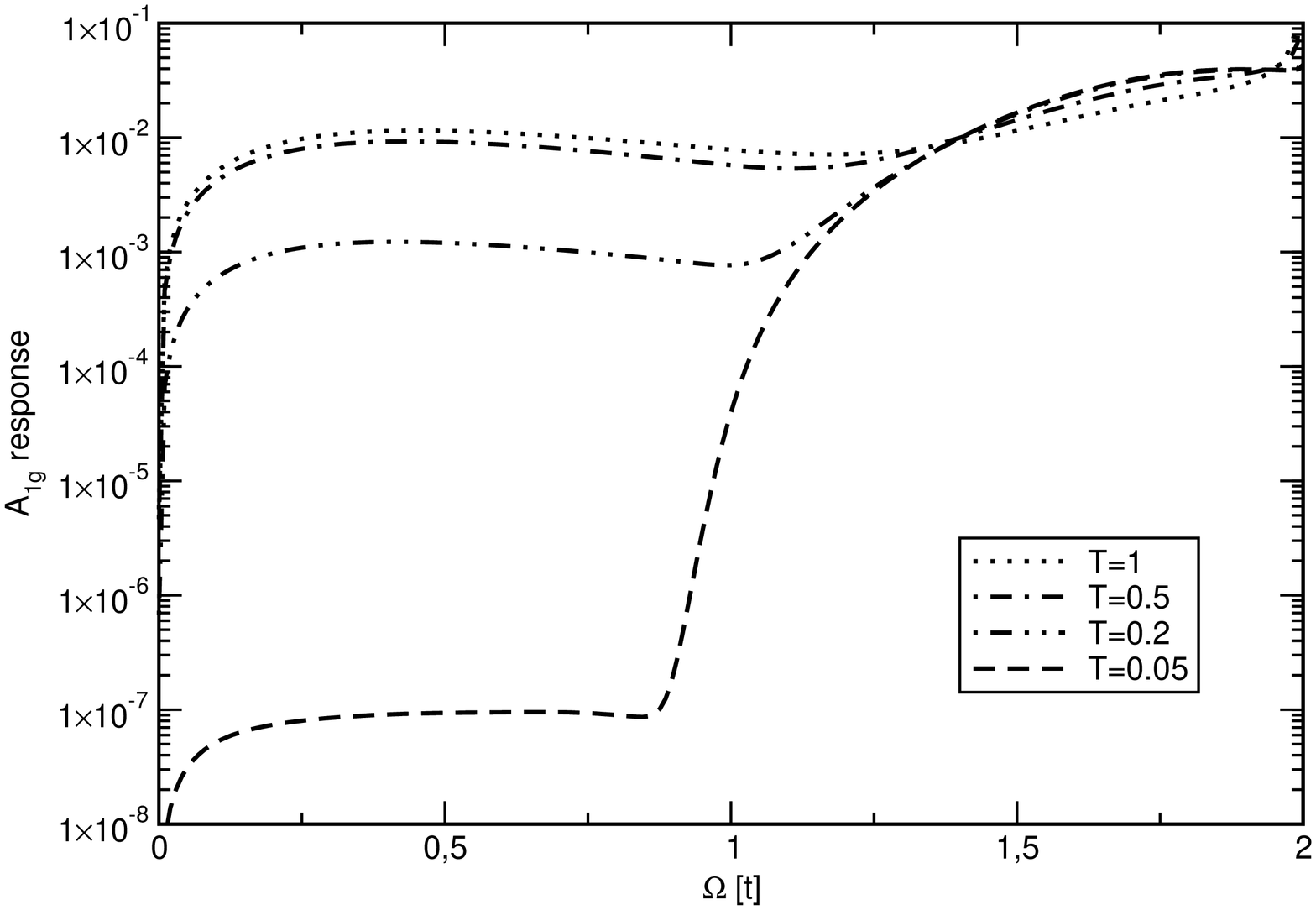}
\caption{The same plot as in Fig.~\ref{fig:raman_iso1}, but for the case $\omega_i=2$. \label{fig:raman_iso2}}
\end{figure}

Our results for the total Raman response as a function of the transfered frequency
$\Omega$ for different temperatures appear in Fig.~\ref{fig:raman_iso1} 
for the incident photon frequencies $\omega_i=2.5$ and $\infty$, and in Fig.~\ref{fig:raman_iso2} for $\omega_i=2$. 

For the $\omega_i=\infty$ case (thin lines in Fig.~\ref{fig:raman_iso1}) we have a pure nonresonant response that is nonzero only in the $A_{\mathrm{1g}}$ and $B_{\mathrm{1g}}$ channels.
One can see, that in both channels all lines, which correspond to different temperatures, cross at a characteristic frequency $\Omega\approx U/2$ (isosbestic point) where the Raman response is independent of temperature. For the Falicov-Kimball model at half-filling, all of the temperature dependence in Eq.~(\ref{chi_bare}) enters from the Fermi distribution function $f(\omega)$. In the insulating phase, the rapidly varying parts of the Fermi distribution functions are located in the gap regions of the two-particle density of states, and the temperature dependence is strongly reduced when these gaps become symmetric for some values of the transfered and incident frequencies.

In Ref.~\cite{paper1}, an isosbestic behavior was observed for the nonresonant response only in the $B_{\mathrm{1g}}$ channel, but in Fig.~\ref{fig:raman_iso1} it is also seen in the $A_{\mathrm{1g}}$ channel when the response is plotted on a logarithmic scale. When the incident photon frequency decreases (thick lines in Fig.~\ref{fig:raman_iso1}), we also have a non-zero Raman response in the $B_{\mathrm{2g}}$ channel and the shape of the Raman response is changed. In particular, a sharp peak appears at the double resonance located at $\Omega=\omega_i$, and the full response is not just an enhancement of the nonresonant features (which are apparent when the incident photon frequency becomes large), but the shape of the response can change dramatically due to resonant effects. This is most apparent when the incident photon energy is close to $U$. Also, the initial $\omega_i=\infty$ isosbestic point is shifted and a second isosbestic point appears in the double resonance area of $\Omega\approx\omega_i$. With a further decrease of the incident photon frequency, both isosbestic points approach one another (Fig.~\ref{fig:raman_iso2}) and then disappear when $\omega_i\approx U/2$.

\begin{figure}
\centering
\includegraphics[height=0.99\columnwidth,angle=-90,clip]{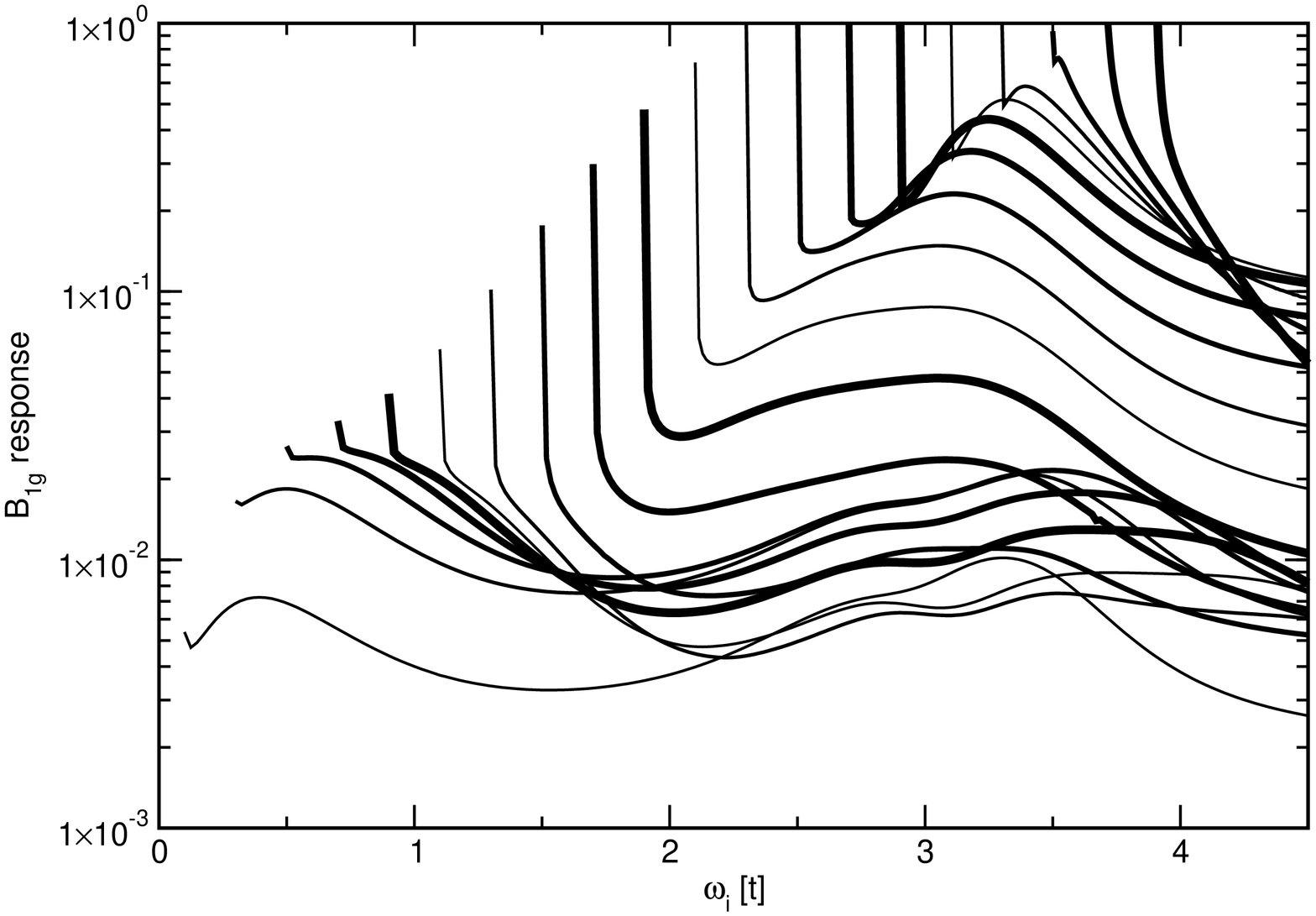}\\
\includegraphics[height=0.99\columnwidth,angle=-90,clip]{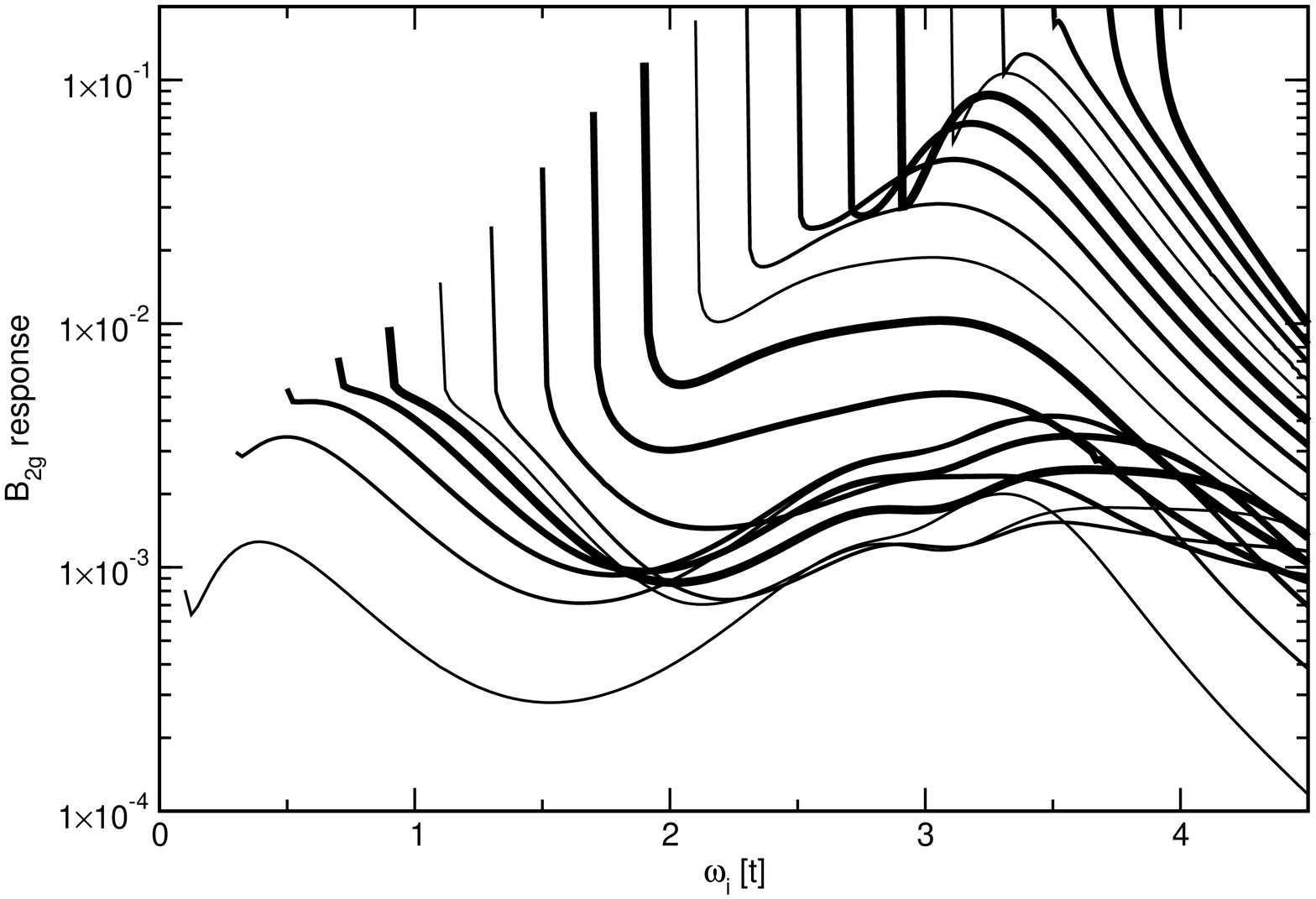}\\
\includegraphics[height=0.99\columnwidth,angle=-90,clip]{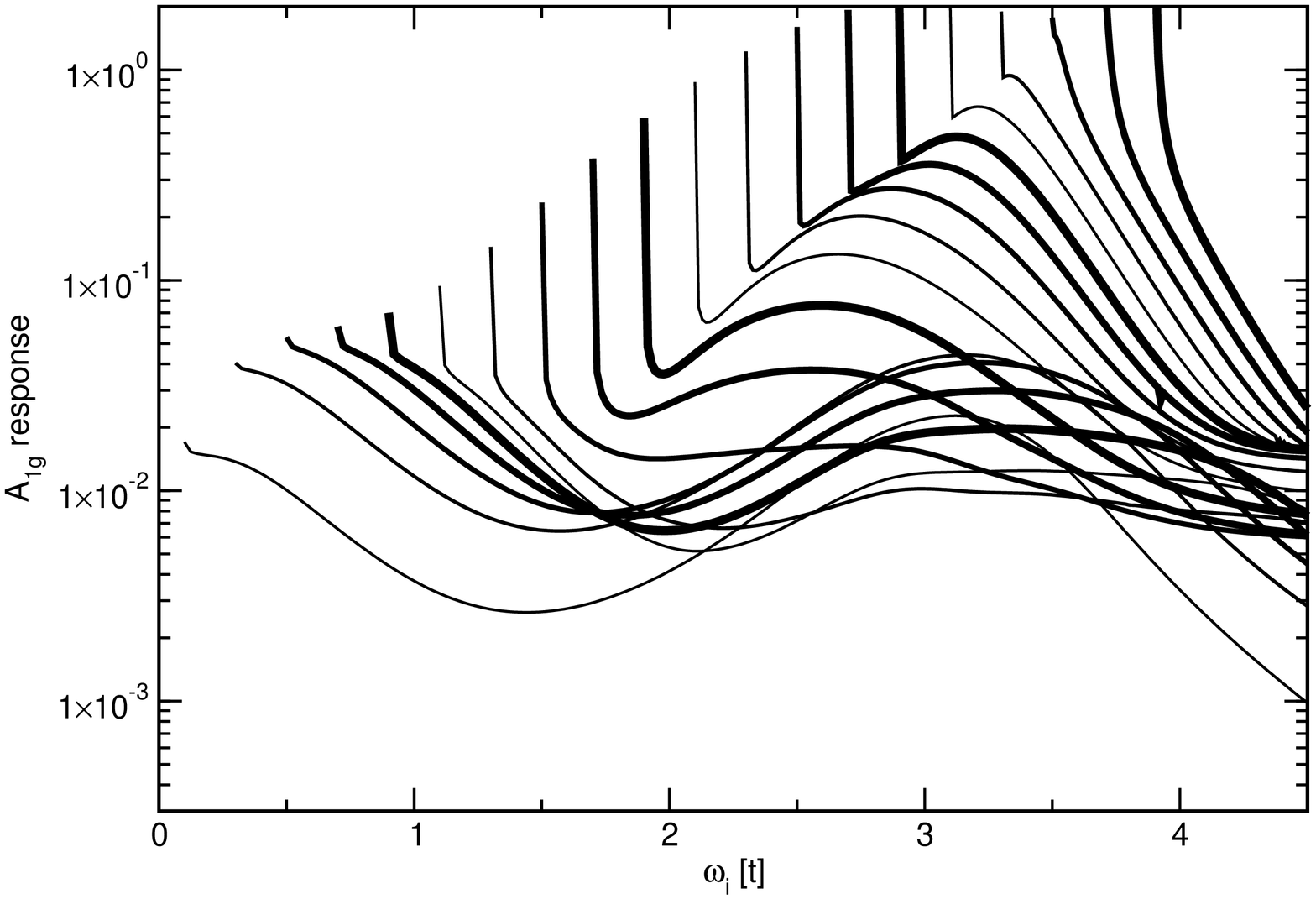}
\caption{Raman response at various values of the transfered frequency $\Omega$ in steps of 0.2 as a function of the incident photon frequency $\omega_i$ for $T=0.5$. All curves start at $\omega_i=\Omega$.
\label{fig:raman_prof}}
\end{figure}

In Fig.~\ref{fig:raman_prof}, we plot the total Raman
profile at various transfered frequencies $\Omega$ as a function of the incident photon frequency $\omega_i$.  Note that when $\Omega$ is larger than the energy of the charge-transfer excitation ($\Omega>U$) we only observe the double resonance at $\omega_i=\Omega$ and there are no additional features in the resonant profile. When $\Omega$ decreases and moves into the charge-transfer peak region, a resonant enhancement of the charge-transfer peak at $\omega_i\approx U$ is observed. Its location and width change with a further decrease of the transfered frequency and it almost disappears when $\Omega$ lies between the charge-transfer and low-energy peaks, and is then restored when $\Omega$ moves into the low-energy peak region, where it has a double-peak structure for the $B_{\mathrm{1g}}$ and $B_{\mathrm{2g}}$ channels. So, we observe a joint resonance of the charge-transfer and low-energy peaks, and this resonance of a low-energy feature due to a higher-energy photon has been seen in the
Raman scattering of some strongly-correlated materials.

In conclusion, we have performed an exact calculation of the electronic Raman response function for a strongly correlated system in the insulating phase and predict four interesting resonant features:
(1) the appearance of a double-resonance peak, (2) the nonuniform enhancement of non-resonant features due to resonance, (3) the appearance of two isosbestic points in all channels, and (4) a joint resonance of the charge transfer and low energy peaks when the incident photon frequency is on the order of $U$.
It will be interesting to see whether these features can be seen in future
experiments on correlated systems.

\textit{Acknowledgments}: We 
acknowledge support from the U.S. Civilian Research
and Development Foundation (CRDF Grant No. UP2-2436-LV-02).
J.K.F. also acknowledges support from the National Science
Foundation under Grant No. DMR-0210717. T.P.D. acknowledges the
NSERC, PREA and the Alexander von Humboldt foundation for support.

\end{document}